\documentstyle[prl,aps,floats,epsf]{revtex}
\begin{document}
\twocolumn[\hsize\textwidth\columnwidth\hsize\csname@twocolumnfalse%
\endcsname

\draft
\title{\Large\bf SU$_M$(2)$\times$U$_C$(1) Gauge Symmetry
in High $T_c$ Superconductivity}

\author{\bf Wei-Min Zhang}

\address{Department of Physics, National Cheng Kung
University, Tainan, 701, Taiwan}

\date{\today}
\maketitle

\begin{abstract}
The square lattice structure of $CuO_2$ layers
and the strongly correlated property of electrons
indicate that the high $T_c$ superconductivity in
cuprates can be described by a SO$_M$(5) coherent
pairing state in which a SU$_M$(2)$\times$U$_C$(1)
gauge symmetry is embedded.
The spin and charge fluctuations that characterize
the low energy magnetic excitations in cuprates are
controlled by this intrinsic SU$_M$(2)$\times$U$_C$(1)
gauge symmetry.
\end{abstract}
%
 ]


The study of low energy quantum fluctuations in
strongly correlated systems is one of the most
difficult problems in theoretical physics. In
copper-oxides, the low energy properties of electrons
are indeed the central issue in the study of high
$T_c$ superconducting mechanism. Up to date, there
exist no proper skills and tools to deal with low
energy quantum fluctuations
of strongly correlated electrons.
However, solving a strongly correlated problem
depends crucially on how to extract the relevant
degrees of freedom that characterize low energy
excitations. Determining low energy degrees
of freedom lies mainly on intrinsic symmetries
of the system. In this letter, I propose that
the low-lying magnetic excitations observed
\cite{41mev,imf,INS} over a large range of dopings
($0.05\leq \delta \leq 0.25$) in cuprates
originate neither from the conventional spin density
wave (SDW) fluctuation associated with the
breaking SU$_S$(2) spin rotational symmetry, nor
from the charge density wave (CDW) fluctuation
related to the breaking U$_C$(1) or SU$_C$(2)
charge gauge symmetry
of singlet pairs. There is an additional intrinsic
symmetry, the SU$_M$(2)$\times$U$_C$(1) gauge
symmetry that mixes the triplet magnetic pairs and
singlet charge pairs.
It is this SU$_M$(2)$\times$U$_C$(1) gauge
symmetry that controls the low energy quantum
fluctuations of spins and charges in cuprates.

In accordance with experimental observations, high $T_c$
superconductivity (SC) arises as a consequence of hole (or
electron) dopings from the parent copper-oxide compounds
which are antiferromagnetic (AF) Mott insulators\cite{dag94}.
In the AF phase which is very closed to the half-filling
($\delta \leq 0.03$), the low-lying excitations are mainly
the SDW fluctuation with respect to the SU$_S$(2) spin
rotational symmetry\cite{Aue}. In the optimal dopings
($0.15 \leq \delta \leq 0.3$), the $d$-wave SC order
\cite{d-wave} implies that the low-lying excitations should
be dominated by phase fluctuation associated with the U$_C$(1)
gauge symmetry, in according to Anderson's RVB theory.
\cite{and87} While in the underdoping region between
the AF and dSC phases, Wen and Lee proposed that there
may exist a SU$_C$(2) gauge symmetry that could address
the pseudogap property\cite{wen96}.

However, the optimally doped YBa$_2$Cu$_3$O$_{6+\delta}$
in SC phase displays a sharp magnetic resonance centered
at $(\pi, \pi)$ in reciprocal space\cite{41mev}.
Meanwhile, in the underdoping region,
especially around $\delta \sim 1/8$, neutron-scattering
experiments on La$_{2-x}$Sr$_x$CuO$_4$ show clear evidences
of incommensurate magnetic excitations disposed
symmetrically about $(\pi, \pi)$\cite{imf}. These
magnetic properties cannot be explained by U$_C$(1)
or SU$_C$(2) charge gauge symmetry. C.S.~Zhang has
argued \cite{szhang97} that the magnetic $\pi$ resonance
may imply the existence of  a SO$_Z$(5) superspin
symmetry\cite{dem95}.
The discovery of incommensurate peaks furthermore brings
another important issue, i.e. possible existence of a stripe
phase\cite{stripe}. More recently, many experiments
have confirmed that incommensurate magnetic excitations
exist not only in LSCO but also in YBCO copper-oxide
superconductors in both underdoping and optimal doping
regions\cite{INS}. From the universality of observed
incommensurate structures, it is natural to ask
whether there exists an intrinsic symmetry to
underlying these low-lying magnetic degrees of
freedom over the whole range of doped cuprates.

An intrinsic symmetry in condensed matter physics
depends not only on basic interacting properties
of electrons, but also on crystal structures of
materials. In general, in terms of the 8-dim
basis of charge, spin and crystal wave vector (in
the reduced first Brillouin zone),
the noninteracting electrons on lattice have a maximum
U(8) symmetry\cite{sol87}. However, for the
high $T_c$ superconductors, the conductivity
is mainly in the $CuO_2$ layers. In such a
two-dimensional square lattice plane,
the parity symmetry between ${\bf k}$ and $-{\bf k}$
reduces the general U(8) group to a smaller subgroup
SO(8)\cite{mar98}. The RVB U$_C$(1) and SU$_C$(2)
gauge symmetries that describe the $d$-wave
superconductivity (SC) and the pseudo-gap behavior
\cite{and87,wen96} and the SO$_Z$(5) superspin
symmetry that describes the antiferromagnetism (AF)
to $d$-wave SC transition \cite{szhang97} are
all subgroups of SO(8).

To realize this SO(8) structure, I introduce a Nambu
basis, $\Psi^\dagger_{\bf k} = (\alpha^\dagger_{\bf k},
\alpha_{\bf -k})$ with $\alpha^\dagger_{\bf k}=(
c^\dagger_{\bf k \uparrow},
c^\dagger_{\bf k \downarrow},c^\dagger_{\bf k+Q \uparrow},
c^\dagger_{\bf k+Q \uparrow})$ and $\alpha_{\bf -k}=(
c_{\bf -k \uparrow}, c_{\bf -k \downarrow},
c_{\bf -k+Q \uparrow}, c_{\bf -k+Q \downarrow})$.
The generators of SO(8) are then given by
\begin{equation}  \label{so8g}
    \Psi^\dagger_{\bf k} {\cal O}_{SO(8)}\Psi_{\bf k}=
    \pmatrix{ \alpha^\dagger_{\bf k}
    {\bf b}_l \alpha_{\bf k} & \alpha^\dagger_{\bf k}
    \bbox{\Delta}^\dagger_i\alpha^\dagger_{\bf -k} \cr
     ~~ & ~~ \cr
    \alpha_{\bf -k}\bbox{\Delta}_i\alpha_{\bf k} &
    -\alpha_{\bf -k}{\bf b}^t_l \alpha^\dagger_{\bf -k}
    \cr}
\end{equation}
where ${\bf k}$ is restricted in the reduced first
Brillouin zone, and ${\bf b}_l$ and its transposition
${\bf b}^t_l$ are $4\times 4$ hermitian matrices,
${\bbox \Delta}_i$ and its hermitian conjugate ${\bbox
\Delta}^\dagger_i$ are $4\times 4$ antisymmetric matrices.
Eq.~(\ref{so8g}) consists of 12 pair operators and 16
particle-hole operators. The 12 pair operators are the
isotropic $s$-wave \cite{BCS} or the $d_{xy}$-wave pair
\cite{dxy} denoted by $\Delta_s$, the extended $s$-wave
or $d_{x^2-y^2}$-wave pair $\Delta_d$ \cite{and87,aff88},
the quasispin $\eta$ pair $\Delta_\eta$ \cite{yang89}
and the three triplet $\pi$ pairs
$\bbox{\Delta}_{\bbox{\pi}}$ \cite{dem95}, plus their
hermitian conjugates. These pair operators are
represented in Eq.~(\ref{so8g})
by $\Delta_{i\bf k} = \alpha_{\bf -k}\bbox{\Delta}_i
\alpha_{\bf k}$ and $\Delta^\dagger_{i\bf k}=
(\Delta_{i\bf k})^\dagger=\alpha^\dagger_{\bf k}
\bbox{\Delta}^\dagger_i\alpha^\dagger_{\bf -k}$, where
$i=s, d, p, {\bbox \pi}$, and $\Delta_i={1\over 2}
(\gamma^x\gamma^z, i\gamma^5\gamma^y,-i\gamma^0\gamma^y,
\gamma^x\gamma^z\bbox{\gamma})$, and the $\gamma$-matrix
is given in the standard Dirac representation in
field theory:
\begin{equation}
    \gamma^0=\pmatrix{ I_2 & 0 \cr 0 & -I_2 \cr }~ , ~~
    \mbox{\boldmath $\gamma$}=\pmatrix{ 0 &
    \mbox{\boldmath $\sigma$} \cr -\mbox{\boldmath
    $\sigma$} & 0 \cr },
\end{equation}
$\gamma^5=i\gamma^0\gamma^x\gamma^y\gamma^z$.
The 16 particle-hole operators include the charge $Q$,
the intrinsic charge $T$, the spin ${\bf S}$,
the intrinsic spin ${\bf A}$, the charge density
wave $C_{\bf Q}$, the charge current $J_c$, the
spin density wave ${\bf S}_{\bf Q}$ and the spin
current ${\bf J}_s$. They are given in Eq.~(\ref{so8g})
by $B_{l\bf k}=\alpha^\dagger_{\bf k}{\bf b}_l
\alpha_{\bf k}-\alpha_{\bf -k}{\bf b}^t_l
\alpha^\dagger_{\bf -k}$ with ${\bf b}_l={1\over 2}
(I_4, \gamma^0,\bbox{\gamma}/2,\gamma^0\bbox{\gamma}/2i,
\gamma^5,i\gamma^5\gamma^0,\bbox{\gamma}\gamma^5/2,
\gamma^5\gamma^0\bbox{\gamma}/2)$ that corresponds to
$B_{l\bf k} \equiv (Q, T, {\bf J}_s, {\bf S}_{\bf Q},
C_{\bf Q}, J_c,  {\bf A}, {\bf S})_{\bf k}$.

Using the generalized coherent state theory
\cite{wzhang90}, a general quasiparticle picture
in terms of the SO(8) Nambu basis can be defined
as follows:
\begin{equation} \label{qp}
    \alpha_{\bf k} | 0 \rangle = 0 ~~~~
        \stackrel{SO(8)}{\longrightarrow} ~~~~
        \beta_{\bf k} |\Omega_{SO(8)} \rangle = 0 .
\end{equation}
The quasiparticle vacuum state $|\Omega_{SO(8)}\rangle$
is given by
\begin{eqnarray}\label{gcs1}
   |\Omega_{SO(8)} \rangle &=&  {\prod}_{\bf k} \exp
   \Big\{\eta_s({\bf k})\Delta^\dagger_{\bf k} +
   \eta_d({\bf k})\Delta^\dagger_{\bf k}  \nonumber \\
& & ~~~ + \eta_p({\bf k}) \Delta_{p\bf k}^\dagger
    +{\bbox \eta}_\pi ({\bf k}) \cdot
    \bbox{\Delta}^\dagger_{\pi \bf k}
    - {\rm H.c} \Big\} | 0 \rangle,
\end{eqnarray}
where the crystal momentum {\bf k} covers the first
Brillouin zone, and $\Delta_{\bf k}^\dagger=
c^\dagger_{{\bf k} \uparrow} c^\dagger_{-{\bf k}
\downarrow}$, $\Delta_{p\bf k}^\dagger=c^\dagger_{
{\bf k} \uparrow}c^\dagger_{-{\bf k+Q}\downarrow}$,
and $\bbox{\Delta}^\dagger_{\pi \bf k} =- c^\dagger_{
{\bf k}\alpha}(i{\bbox \sigma}\sigma^y)_{\alpha\beta}
c^\dagger_{-{\bf k+Q}\beta}$ are various singlet
and triplet $\pi$ pair operators in the SO(8) algebra.
The quasiparticle
operators $(\beta^\dagger_{\bf k}, \beta_{\bf k})$
are determined by the corresponding SO(8)
Bogoliubov transformation with respect to the physical
vacuum state $|\Omega_{SO(8)}\rangle$ \cite{wzhang20a},
\begin{equation} \label{bt}
\pmatrix{ \beta_{\bf k} \cr \beta^\dagger_{\bf -k}
    \cr} = \pmatrix{ W_{\bf k} & -Z_{\bf k} \cr
    Z^\dagger_{\bf k} & W^t_{\bf k} \cr} \pmatrix{
    \alpha_{\bf k}\cr \alpha^\dagger_{\bf -k} \cr},
\end{equation}
The notation $Z_{\bf k}$ and $W_{\bf k}$ are $4\times
4$ block matrices in the canonical SO(8) Bogoliubov
transformations:
\begin{equation} \label{btm}
    Z_{\bf k}= \eta {\sin\sqrt{\eta^\dagger
    \eta} \over \sqrt{\eta^\dagger \eta}}, ~~~
    W_{\bf k}=\cos\sqrt{\eta \eta^\dagger}
\end{equation}
and
\begin{equation} \label{etm}
    \eta =\pmatrix{0 & \eta_1({\bf k}) &
    \eta_2({\bf k}) & \eta_3({\bf k}) \cr
    -\eta_1({\bf k}) & 0 & \eta_4({\bf k})
    & \eta_5({\bf k}) \cr -\eta_2({\bf k})&
    -\eta_4({\bf k}) & 0 & \eta_6({\bf k}) \cr
    -\eta_3({\bf k}) & -\eta_5({\bf k}) &
    -\eta_6({\bf k}) & 0 \cr}
\end{equation}
with $\eta_{1,6}({\bf k})=\eta_s({\bf k})\pm
\eta_d({\bf k})$, $\eta_{4,3}({\bf k})=
\eta^z_{\pi}({\bf k})\pm \eta_p({\bf k})$,
$\eta_2({\bf k})=\eta^-_{\pi}({\bf k})$
and $\eta_5({\bf k})=-\eta^+_{\pi}({\bf k})$.

The state
$|\Omega_{SO(8)} \rangle$ is nothing but the SO(8)/U$_g$(4)
coherent pairing state which is, apart from a phase
factor, obtained by acting SO(8) on the trivial
vacuum $|0\rangle$\cite{wzhang87}, while the associated
phase factor contains the freedom of U$_g$(4) gauge
transformations that describe quantum fluctuations
of all the pairing wave functions $\eta_i({\bf k})$.
The pairing wave functions $\eta_i({\bf k})$ in
Eq.~(\ref{gcs1}) are generally {\it link-dependent}
complex parameters with an additional
constraint $\eta_i({\bf k}) = \eta_i({\bf -k})$
from the parity symmetry.
In fact, $|\Omega_{SO(8)}\rangle$ is the underlying pairing
state I proposed recently to describe high $T_c$
superconductivity \cite{wzhang20}. As I have discussed
in \cite{wzhang20}, $|\Omega_{SO(8)}\rangle$ consists of
all electron pairs concerned in the study of
superconductivity. These pairs can be classified
according to the symmetric property of the pairing
wave function under the transformation of ${\bf k}$
to ${\bf k+Q}$ as follows: $\eta_s({\bf k}) =
\eta_s({\bf k+Q})$ represents the isotopic
$s$-wave and $d_{xy}$-wave ($\sim \sin k_x \sin k_y$)
singlet pairs etc., $\eta_d({\bf k})=-\eta_d({\bf k+Q})$
describes the extended singlet pairs [including
the extended $s$-wave ($\sim \gamma({\bf k})=\cos k_x
+ \cos k_y$), the $d$-wave ($\sim d({\bf k})=\cos k_x
-\cos k_y$) and the $s+id$-wave ($\sim \cos k_x
+i\cos k_y$) pairs etc.], while $\eta_p({\bf k})
=\eta_p({\bf k+Q})$ corresponds to the pseudo-spin
pairs, and finally, ${\bbox \eta}_\pi ({\bf k})=-{\bbox
\eta}_\pi({\bf k+Q})$ describe the triplet $\pi$ pairs.

Since high $T_c$ superconductors are obtained
by doping from the parent copper-oxides which
are AF insulators, to demonstrate how can this
picture be realized and what are the low-lying
degrees of freedom involved, I should first check
the AF order parameter and the hopping dynamics
contained in the state $|\Omega_{SO(8)}\rangle$. Without loss
generality, I define ${\bbox \eta}_\pi({\bf k})
= \eta_\pi({\bf k}){\bbox \alpha}_{\bf k}$ with
${\bbox \alpha}_{\bf k}$ being a unit vector,
$|\bbox{\alpha}_{\bf k}| =1$.
Then the AF order in $|\Omega_{SO(8)}\rangle$ is given by
\begin{eqnarray} \label{af}
{\bf m}_{AF} &\equiv& {1\over N} \langle \Omega_{SO(8)} |
    {1\over 2}\sum_{\bf k}c^\dagger_{\bf k\alpha}
    {\bbox \sigma}_{\alpha\beta}c_{\bf k+Q\beta}
    |\Omega_{SO(8)}\rangle \nonumber \\
&=& {2\over N}{\sum}'_{\bf k}[z_d({\bf k})z^*_\pi({\bf k})
    +z^*_d({\bf k}) z_\pi({\bf k})]{\bbox \alpha}_{\bf k} ,
\end{eqnarray}
where  $z_i({\bf k})$
is an element of $Z_{\bf k}$ in Eq.~(\ref{btm}), which is
defined in the same form as in Eq.~(\ref{etm}).
For the leading hopping Hamiltonian, one can obtain
\begin{eqnarray} \label{hh}
\langle H_t \rangle
    &=& \langle \Omega_{SO(8)} |\sum_{\bf k \sigma}
    \varepsilon({\bf k})c^\dagger_{\bf k\sigma}
    c_{\bf k\sigma} | \Omega_{SO(8)} \rangle  \nonumber \\
    &=& 4 {\sum}'_{\bf k} \varepsilon({\bf k})[z_s({\bf k})
    z^*_d({\bf k}) + z^*_s({\bf k})z_d({\bf k})] \label{tterm}
\end{eqnarray}
and $\varepsilon({\bf k}) = -2t (\cos k_x + \cos k_y)$.
Eq.~(\ref{af}) simply tells us that the AF state mixes
the extended singlet pairs and the triplet $\pi$ pairs.
Meanwhile, Eq.~(\ref{hh}) shows that the hoping requires
the simultaneously present of the sinlget pairs of
$\eta_s({\bf k})$ and $\eta_d({\bf k})$.

Zhang's SO$_Z$(5)
superspin theory\cite{szhang97} contains only the $d$-wave
singlet pair plus the triplet $\pi$ pairs. The corresponding
pairing state can be generally expressed as
\begin{eqnarray}
   &&|\Omega_{SO(6)} \rangle = {\prod}_{\bf k} \exp
    \Big\{\eta_d({\bf k})\Delta^\dagger_{\bf k}+
    {\bbox \eta}_\pi ({\bf k}) \cdot \bbox{\Delta}^\dagger_{
    \pi \bf k} - {\rm H.c} \Big\} | 0 \rangle
    \nonumber \\ & & ~~~  \nonumber \\
   &&~~~= \left\{ \begin{array}{l}{\prod}_{\bf k} \exp \Big\{
    {\bbox \eta}_\pi ({\bf k}) \cdot \bbox{\Delta}^\dagger_{
    \pi \bf k} - {\rm H.c} \Big\} | AF \rangle \\
    {\rm or } \\
    {\prod}_{\bf k} \exp \Big\{{\bbox \eta}_\pi ({\bf k})
    \cdot \bbox{\Delta}^\dagger_{\pi \bf k} - {\rm H.c}
    \Big\} | \begin{array}{c}d{\rm -wave} \\ {\rm BCS} \\
    \end{array} \rangle  \end{array} \right. , \label{gcs2}
\end{eqnarray}
which gives a realization of the picture how the
$\pi$-operator rotates the AF state into $d$-wave SC and
vis-versus.  However, from Eq.~(\ref{hh}), one sees
that the SO$_Z$(5) theory cannot describe hopping
dynamics because of the lacking of the $s$-wave
(or $d_{xy}$-wave) type pairing. This is why the doping
process is artificially addressed through a chemical
potential in Zhang's SO$_Z$(5) theory.
On the other hand, Anderson's RVB state only consists
of the extended singlet pairs,
\begin{equation}\label{gcs3}
   |\Omega_{RVB} \rangle = P_G{\prod}_{\bf k} \exp \Big\{
    \eta_d({\bf k})\Delta^\dagger_{\bf k}- {\rm H.c}
    \Big\} | 0 \rangle ,
\end{equation}
so that it cannot describe the AF magnetization,
where $P_G$ is the Gutzwiller projector that removes the
doubly occupied sites when it is applied to the $t-J$
model \cite{and87}. Consquently, both Anderson's RVB
theory and Zhang's SO$_Z$(5) theory only catch a partial
low-lying degree of freedom in high $T_c$ superconductivity.

On the other hand, the unit vector ${\bbox \alpha}_{\bf k}$
in (\ref{af}) is the normalized N\'{e}el field that
describes a continuous manifold of degenerate ground
states in $|\Omega_{SO(8)}\rangle$ with regard to the SU$_S$(2)
spin rotational symmetry. To highlight the dominated
low-lying degrees of freedom in cuprates, I can rewrite
Eqs.~(\ref{gcs1}) equivalently as
\begin{equation}
|\Omega_{SO(8)} \rangle \rightarrow \exp\{{\sum}_{\bf k}{\bbox
    \theta}({\bf k}) \cdot{\bf S}_{\bf k}\}|\Omega_{SU(4)}
    \rangle,
\end{equation}
where $|\Omega_{SU(4)}\rangle$ is defined as an
intrinsic state in which ${\bbox
\alpha}_{\bf k}$ has been specified to along the easy
$z$-axis:
\begin{eqnarray}\label{gcs4}
   |\Omega_{SU(4)} \rangle = && {\prod}_{\bf k} \exp
        \Big\{\eta_s({\bf k}) \Delta^\dagger_{\bf k} +
        \eta_d({\bf k})\Delta^\dagger_{\bf k}  \nonumber \\
& & ~~~~ + \eta_p({\bf k}) \Delta_{p\bf k}^\dagger
        + \eta^z_\pi ({\bf k}) \Delta^\dagger_{\pi^z \bf k}
        - {\rm H.c} \Big\} | 0 \rangle .
\end{eqnarray}
In fact, the intrinsic pairing state
$|\Omega_{SU(4)}\rangle$ is a reduction of the
SO(8)/U$_g$(4) to
SU(4)/SU$_C$(2)$\times$SU$_M$(2)$\times$U$_C$(1)
coherent pairing state under the spin rotational
symmetry. Under this decomposition,
the conventional SDW arisen from spin fluctuation
can be determined by varying ${\bbox \theta}_{\bf k}$.
It can be shown that both
the hoping [see Eq.~(\ref{hh}] and the doping,
\begin{eqnarray}
\delta &=& \langle \Omega_{SO(8)} |1-
    {\hat{n}\over N} | \Omega_{SO(8)} \rangle \nonumber\\
    &=& 4{\sum}_{\bf k}[1-|z_s({\bf k})|^2 +|z_d({\bf k})|^2
    +|z_p({\bf k})|^2 + |z_\pi({\bf k})|^2].
    \label{dopi} \nonumber
\end{eqnarray}
are independent from  ${\bbox
\alpha}_{\bf k}$ [and ${\bbox \theta}({\bf k})$].
In other words, the conventional SDW does not
explicitly depend on dopings.
I thereby conclude that the incommensurability
of the observed magnetic excitations that linearly
depend on dopings\cite{INS} should not originate from
the spin fluctuation of the {\it conventional} SDW.
In fact, it is not the unit spin vector ${\bbox
\alpha}_{\bf k}$ but the amplitude of
staggered AF order,
\begin{eqnarray}  \label{mga}
    m_{AF} &=&{2\over N}{\sum}'_{\bf k}[z_d({\bf k})
    z^*_\pi({\bf k})+z^*_d({\bf k})z_\pi({\bf k})]
\end{eqnarray}
that sensitively depends on
dopings. The low-lying magnetic excitations in cuprates
must arise from quantum fluctuations of the AF amplitude
$m_{AF}$ due to dopings rather than the SDW of spin
fluctuations.

Furthermore,
if I further drop the $z$-component of the $\pi$ pairs,
the state $|\Omega_{SU(4)}\rangle$ is deduced to a
SO$_C$(5)/SU$_C$(2)$\times$U$_C$(1) coherent
pairing state:
\begin{eqnarray}\label{gcs5}
   |\Omega_{SO_C(5)} \rangle = && {\prod}_{\bf k} \exp
        \Big\{\eta_s({\bf k}) \Delta^\dagger_{\bf k} +
        \eta_d({\bf k})\Delta^\dagger_{\bf k}  \nonumber \\
& & ~~~~~~~~~~~~~~~ + \eta_p({\bf k})
    \Delta_{p\bf k}^\dagger - {\rm H.c} \Big\} | 0 \rangle ,
\end{eqnarray}
where the SO$_C$(5) group is obviously not the
SO$_Z$(5) superspin symmetry in Zhang's theory.
The subgroup SU$_C$(2)$\times$U$_C$(1) (generated by
$\{C_Q, J_c, T, Q\}$) is a gauge symmetry embedded
in $|\Omega_{SO_C(5)} \rangle$ that describes
quantum fluctuations of the $s$, $d$-wave pair
orders, the CDW order and staggered flux order.
The CDW and staggered flux orders
are given by
\begin{eqnarray} \label{cd}
    \rho_{\bf Q} &\equiv & {1\over N} \langle \Omega_{SO(8)} |
    \sum_{\bf k \sigma}c^\dagger_{\bf k \sigma}
    c_{\bf k+Q \sigma}|\Omega_{SO(8)} \rangle \nonumber \\
    &=& {4\over N}{\sum}'_{\bf k}[z_s({\bf k})
    z^*_p({\bf k})+z^*_s({\bf k}) z_p({\bf k})] , \\
    \chi &\equiv& {1\over N} \langle \Omega_{SO(8)}
    |{\sum}_{<ij>\sigma}c^\dagger_{i\sigma}c_{j\sigma}
    |\Omega_{SO(8)} \rangle \nonumber \\
    &=& {4\over N}{\sum}'_{\bf k}\gamma({\bf k})[z_p({\bf k})
    z^*_d({\bf k})+ z^*_p({\bf k})z_d({\bf k})] ,
\end{eqnarray}
respectively. It covers
the SU$_C$(2) RVB gauge theory proposed by Wen and Lee when
one imposes the constraint of no-double-occupied sites
on the state $|\Omega_{SO_C(5)} \rangle$ \cite{wen96}.
This state is capable of describing the coexistence of the
$d$-wave superconductivity with the CDW and staggered
flux phases. But it cannot address the dynamics of
AF magnetic phase.

Upon date, there is no direct evidence
on the existence of long range CDW or staggered
flux order in cuprates. To concentrate on the observed
magnetic excitations, I may set $\eta_p({\bf k})=0$ in
Eq.~(\ref{gcs4}).
Then, the intrinsic pairing state $|\Omega_{SU(4)}\rangle$
is reduced to another SO(5) coherent state, specifically,
the SO$_M$(5)/SU$_M$(2)$\times$U$_C$(1) coherent pairing state,
\begin{eqnarray}\label{gcs6}
   |\Omega_{SO_M(5)} \rangle = && {\prod}_{\bf k} \exp
        \Big\{\eta_s({\bf k}) \Delta^\dagger_{\bf k} +
        \eta_d({\bf k})\Delta^\dagger_{\bf k}  \nonumber \\
& & ~~~~~~~~~~~~~~
        + \eta^z_\pi ({\bf k}) \Delta^\dagger_{\pi^z \bf k}
        - {\rm H.c} \Big\} | 0 \rangle .
\end{eqnarray}
From Eqs.~(\ref{hh}) and (\ref{mga}), one sees that
this SO$_M$(5) coherent pairing state can describe the
hoping dynamics and the coexistence of the $s$,
$d$-wave superconductivities
and the AF magnetism with local magnetic excitations.
While, the subgroup SU$_M$(2)$\times$U$_C$(1) (generated
by $\{S^z_{\bf Q},J^z_s,T,Q\}$) is the gauge symmetry
embedded in $|\Omega_{SO_M(5)} \rangle$ that controls
quantum fluctuations of the $s$, $d$ and $\pi^z$
pairs. The physical properties of $|\Omega_{SO_M(5)}
\rangle$ in Eq.~(\ref{gcs6}) are very different from
 $|\Omega_{SO_C(5)} \rangle$ in Eq.~(\ref{gcs5}).
The state $|\Omega_{SO_M(5)} \rangle$ is magnetic
charge dominated, while $|\Omega_{SO_C(5)} \rangle$ is
charge dominated. It is the former that can effectively
describe the low-lying magnetic excitations.

In conclusion, the above analysis shows that the low energy
degrees of freedom in cuprates that compasses SDW, CDW,
staggered flux order, magnetic excitations and
various pairing (including the $s$, $d$-wave
singlet and $\pi$ triplet pair) orders
can be encompassed by the SO(8)/U(4) coherent pairing
state $|\Omega_{SO(8)}\rangle$ [i.e. Eq.~(\ref{gcs1})].
The SO(8) coherent pairing state
contains three different SO(5) subgroup pairing
states. The first two states, Eqs.~(\ref{gcs3}) and
(\ref{gcs5}), cover Zhang's SO$_Z$(5) theory
and Wen-Lee's SU$_C$(2) gauge theory, respectively,
and the last one, the SO$_M$(5) symmetry is a new
discovery.
\begin{equation}
    |\Omega_{SO(8)}\rangle \rightarrow \left\{
        \begin{array}{l} |\Omega_{SO(6)}\rangle
        \rightarrow |\Omega_{SO_Z(5)}\rangle \\ \\
        |\Omega_{SU(4)}\rangle \rightarrow \left\{
        \begin{array}{c} |\Omega_{SO_C(5)}\rangle \\ \\
        |\Omega_{SO_M(5)}\rangle \end{array}\right.
        \end{array} \right. .
\end{equation}
The above three SO(5) coherent
pairing states are generated
by different pair operators, and they describe different
physical properties of strongly correlated electrons in
cuprates. Only the SO$_M$(5) symmetry is capable of
describing the low-lying magnetic excitations incorporating
with the hoping dynamics.
Specifically, all the three pairing states
contains the $d$-wave superconducting order. However,
they carry different gauge degrees of freedom associated
with different quantum fluctuations:
\begin{equation}
{SO(8)\over U(4)} \rightarrow \left\{
    \begin{array}{l} {SO(6)\over SU_S(2)\times
    SU_{S_{\bf Q}}(2)\times U(1)} \rightarrow
    {SO_Z(5)\over SU_S(2) \times U(1)}\\ \\
    {SU(4)\over SU_C(2)\times SU_M(2)\times U(1)}
    \rightarrow \left\{\begin{array}{c} {SO_C(5)\over
    SU_C(2)\times U(1)} \\ \\ {SO_M(5)\over SU_M(2)
    \times U(1)} \end{array}\right. \end{array} \right.
\end{equation}
In Zhang's SO$_Z$(5) theory, the gauge symmetry is
represented by the spin rotational SU$_S$(2) group
plus the charge U$_C$(1) group. It separately describes
the SDW quantum fluctuation and the U$_C$(1) charge
fluctuation. However, they are not essential to the
magnetic excitations. Also the doping can only be added
artificially through a chemical potential in this theory.
In Wen and Lee's SU$_C$(2) gauge theory, the SU$_C$(2)
gauge symmetry describes the CDW and the staggered flux
phase but no AF feature. Only in the SO$_M$(5) coherent
pairing theory, the SU$_M$(2)$\times$U$_C$(1)
gauge group can simultaneously and {\it dynamically}
addresses quantum fluctuations of the AF amplitude
and hopings as well as the $d$-wave pairing.
This SU$_M$(2)$\times$U$_C$(1) gauge symmetry has
not been realized in the previous study of high $T_c$
theories. It should be this SU$_M$(2)$\times$U$_C$(1)
gauge symmetry
that describes the various magnetic excitations
in cuprates. I will discuss its applications in more
details in a separate publication. This work is supported
by the grant NSC89-2112-M006-029.


\begin{thebibliography}{9}
\vspace{-0.5in}
    \bibitem{41mev} J. Rossa-Mignod et al., {\it Physica}
            (Amsterdam) {\bf 169B}, 58 (1991);
        J. M. Mook et al., {\it Phys. Rev. Lett.}
            {\bf 70}, 3490 (1993);
        H. F. Fong et al. {\it Phys. Rev. Lett.}
            {\bf 75}, 316 (1995).
    \bibitem{imf} S. W. Cheong et al., {\it Phys. Rev. Lett.}
        {\bf 67}, 1791 (1991);
        J. M. Tranquada et al., {\it Phys. Rev. Lett.}
        {\bf 73}, 1003 (1994).
    \bibitem{INS} 
        G. Aeppli et al., {\it Science} {\bf 278}, 1432 (1997);
        K. Yamada et al., {\it Phys. Rev.} {\bf B 57}, 6165 (1998);
        P. Dai et al., {\it Phys. Rev. Lett.} {\bf 80}, 1738 (1998);
        H. A. Mook et al, {\it Nature} {\bf 395}, 580 (1998);
        M. Arai et al., {\it Phys. Rev. Lett.} {\bf 83}, 608 (1999);
        B. Lake et al. {\it Nature} {\bf 400}, 43 (1999);
        F. Ronning et al., {\it Science} {\bf 282}, 2067 (1998);
   \bibitem{dag94} E. Dagotto, {\it Rev. Mod. Phys.}
            {\bf 66}, 763 (1994).
   \bibitem{Aue} A. Auerbach, ``Interacting Electrons and
            Quantum Magnetism'', (Springer-Verlag, 1994).
   \bibitem{d-wave} Z. X. Shen et al., {\it Phys. Rev. Lett.}
            {\bf 70}, 1553 (1993);
        C. C. Tsuei et al., {\it Science} {\bf 271}, 329 (1996).
    \bibitem{and87} P. W. Anderson, {\it Science}
            {\bf 235}, 1196 (1987).
    \bibitem{wen96} X. G. Wen and P. A. Lee, {\it Phys. Rev. Lett.}
        76, 503 (1996).
    \bibitem{szhang97} S. C. Zhang, {\it Science}
            {\bf 275}, 1089 (1997).
    \bibitem{dem95} E. Demler and S. C. Zhang, {\it Phys. Rev. Lett.}
            {\bf 75}, 4126 (1995); {\it Nature} {\bf 369}, 733 (1998).
    \bibitem{stripe} J. M. Tranquada et al., {\it Nature}
        {\bf 375}, 561 (1995). V. J. Emery and S. A. Kivelson,
        cond-mat/9808083.
    \bibitem{sol87} A. I. Solomon, and J. L. Birman,
        {\it J. Math. Phys.} {\bf 28}, 1526 (1987);
        S. \"{O}stlund, {\it Phys. Rev. Lett.}
        {\bf 69}, 1695 (1992).
    \bibitem{mar98} A similar SO(8) algebra based on Van
        Hove singularities has also been constructed by
        R. S. Markiewicz and M. T. Vaughn,
        {\it Phys. Rev.} {\bf B 57}, R14052 (1998).
    \bibitem{BCS} J. Bardeen, L. N. Cooper and J. R. Schrieffer,
        {\it Phys. Rev.} {\bf 108}, 1175(1957).
    \bibitem{dxy} R. B. Laughlin, {\it Phys. Rev. Lett.}
        {\bf 80}, 5188 (1998).
    \bibitem{aff88} I. Affleck, Z. Zou, T. Hsu and P. W. Anderson,
        {\it Phys. Rev.} {\bf B 38}, 745 (1988).
    \bibitem{yang89} C. N. Yang, {\it Phys. Rev. Lett.}
        {\bf 63}, 2144 (1989);
        C. S. Zhang, {\it Phys. Rev. Lett.} {\bf 65}, 120 (1990).
    \bibitem{wzhang90} W. M. Zhang et al., {\it Rev. Mod. Phys.}
        {\bf 62}, 867 (1990).
    \bibitem{wzhang87} W. M. Zhang et al.,
        {\it Phys. Rev. Lett.} {\bf 59}, 2032 (1987).
    \bibitem{wzhang20a} W. M. Zhang, {\it Chin. J. Phys.}
        {\bf 38}, 371 (2000).
    \bibitem{wzhang20} W. M. Zhang, {\it Phys. Rev.}
        {\bf B 61}, R898 (2000).
\end{thebibliography}
\end{document}